# A Simulation-based Optimization Approach to Efficiently Route Air Taxis in a Cyber-Physical Network

Suchithra Rajendran[a,b,*]

[a]Department of Industrial and Manufacturing Systems Engineering, University of Missouri Columbia, MO 65211, USA
[b]Department of Marketing, University of Missouri Columbia, MO 65211, USA

[*]Corresponding Author:
Suchithra Rajendran
E-mail address: RajendranS@missouri.edu
Telephone: 573-882-7421

# A Simulation-based Optimization Approach to Efficiently Route Air Taxis in a Cyber-Physical Network

## Abstract

Besides air pollution and commuter stress, traffic congestions also lead to loss of productivity, increase in delay, vehicle operating cost, and accidents. To assuage these issues, several logistics companies are planning to launch air taxis, electric-powered vehicles that aim to provide faster passenger commutes on a daily basis at an affordable cost. This research is one of the first to propose a centralized framework to dispatch and route flying taxis in a cyber-physical network considering unique constraints pertaining to air taxi operations. The feasibility of the proposed approach is tested using potential air taxi demands in New York City (NYC) provided by a prior study. The results of the experimentation suggest that the minimum number of air taxis required for efficient operation in NYC is 84, functioning with an average utilization rate of 66%. In addition, the impacts of commuter's "willingness to fly" rate, percentage of demand fulfillment, on-road travel limit, maximum customer wait time, and arrival distribution on the optimal number of air taxis, utilization rate, number of customers served and cost incurred per customer are examined. Analyses show that the "willingness to fly" rate appears to have a linear influence on the number of air taxis and the efficiency, while on-road travel distance has an exponential impact on the performance measures. The routing and dispatching algorithm developed in this paper can be used by any company that is interested in venturing into the air taxi market.

# 1. Introduction

Traffic congestion in metropolitan cities not only leads to accidents, stress, and vehicle operating cost, but is also a major cause of air pollution. Additionally, heavy traffic engenders the loss in economic output, jobs, and revenues for businesses across different sectors. For instance, nearly four million commuters travel to work each day to the Manhattan district of New York City (NYC) as this borough is built on almost a trillion-dollar economy (Partnership for New York City, 2019). The inevitable traffic conditions economically affect Manhattan by nearly $13 billion annually with $2 billion wasted only in fuel and vehicle operating cost, and loss of about 45,000 jobs, as depicted in Figure 1 (Partnership for New York City, 2019). Similar issues exist in several other large cities across the globe, such as London, Paris, and Tokyo, as well. Despite the government actively encouraging commuters to carpool, use subways and sometimes even impose a user fee to travel through the central business district, traffic continues to increase due to rapid urbanization and growing businesses.

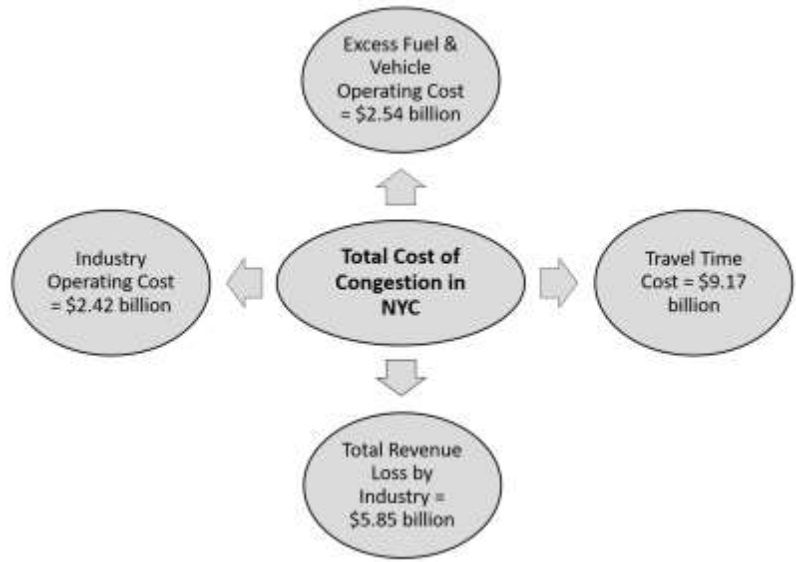

**Figure 1:** Cost of Excess Congestion in the New York Metro Region

Air taxis, electric-powered vehicles, are one of the recent innovations that companies (such as Uber, Zephyr Airworks and Airbus) are planning to launch into metropolitan cities in an attempt to moderate current traffic issues and provide faster passenger commutes (Rajendran and Zack, 2019; Rath and Chow, 2019). Though these flying vehicles cannot serve all passengers commuting on the road every day, millions of riders who are making cumbersome, time-consuming trips are anticipated to benefit from such services. Similar to helicopters, air taxis vertically take-off and land (referred to as the VTOL technology) on architecture that can be retrofitted on rooftops or parking garage (Johnson et al., 2018). This particular feature appears to be an advantage for passengers who will be using these services on a daily basis, and hence the pickup and drop-off locations have to be easily accessible for ensuring shorter ride times and more comfortable mobility.

Several studies have discussed the two types of physical hubs for air taxi network operations (Smith, 1994; Patnoe, 2018; Hasan, 2019). The first type of station is referred to as a vertistop, a single-pad infrastructure (referred to as *vertistop*) in which air taxis can pick up and/or drop customers and take off quickly afterward. The other is a larger facility (referred to as *vertiport*) that can accommodate multiple air taxis takeoff/landing. Air taxis can also halt at vertiports for getting charged, repaired, or when they are idle (Holden and Geol, 2016). Thus, a challenging problem that arises from a logistics point of view is that, when an air taxi drops a customer at a station, an immediate decision has to be made on whether it has to be directed to another facility (where there exist demand for air taxis) or be idle at the nearest vertiport.

The introduction of air taxis will result in several benefits and challenges to the passengers as well as the companies operating them. While commuters are expected to save a significant amount of time using this service, they would have to account for the relatively more expensive travel cost, and the difficulty of going through multi-modal transportation (passengers have to first travel from their locations to the nearest vertistop/vertiport by walk or car and then avail the air taxi service). On the other hand, logistics companies offering these services will have to make several real-time decisions such as (i) evaluating several candidate trips and selecting the schedule that aims to reduce passenger ride time and cost, (ii) coordinating air taxis while landing, taking-off and tracking vehicles on air, (iii) optimizing flight operations by effectively and efficiently routing air taxis to achieve maximum serviceability, (iv) estimating the air taxi market demand based on the geographical region, time of the day and day of the week, and (v) managing battery status and other maintenance-related issues.

The objective of this research is to develop an algorithm that can help logistics companies to dispatch and route air taxis effectively throughout the physical network of vertiports and vertistops. To accomplish this objective, a simulation-based optimization technique is proposed, in which the simulation model interfaces with the optimization solver to provide optimal routing in real-time, considering several unique constraints that are pertaining to air taxi operations. After each customer drop-off, decisions will be made on (i) whether the air taxi must become idle or pick up customers from other sites, and (ii) which station should the air taxi be routed to (if the air taxi is not idle). The proposed model is tested with the estimated air taxi demand data from a prior study (Rajendran and Zack, 2019).

The remainder of the paper is organized as follows. The review of the literature on air taxis and similar studies are presented in Section 2. Data is detailed in Section 3, and the proposed simulation-based optimization algorithm is explained in Section 4. Results and sensitivity analysis are discussed in Section 5. Conclusions and future works are given in Section 6.

## 2. Literature Review

This section presents a review of the literature on air taxis and other similar transportation methods, such as traditional taxi and bike-sharing services.

### 2.1 Air Taxi

Baik et al. (2008) were one of the first to conduct studies pertaining to air taxi operations. The authors developed a mode-choice model in which different modes of transportations were evaluated for round-trip demands. While air taxis were not used for everyday passenger commutes about a decade back, their work is one of the preliminary research in the air taxi design and operations domain. A large number of existing studies in the air taxi arena have examined the vehicle design. Falck et al. (2018) performed an acoustic analysis to reduce the sound pressure level perceived in densely populated areas. Johnson et al. (2018) studied the emerging aviation market to guide NASA's research activities for air taxi development. The authors categorized a wide range of attributes that the flying vehicle design community considers, few of them include the number of passengers, battery range, aircraft type, and propulsion system.

Due to increasing traffic that is encountered globally, air taxi prototypes and testing are conducted in several countries. For instance, Uber has laid its test grounds on metropolitan cities such as Dallas, Los Angeles, as well as planning to perform flight tests internationally in Dubai, Tokyo, Singapore, London, and Bangalore (Hawkins, 2018). Headed by Sebastian Thrun and supported by Google co-founder Larry Page, Kitty Hawk announced in March 2018 the intention to launch its soaring air taxi services (called *Cora*) in New Zealand (Warwick, 2018). Airbus's Vahana, self-piloted eVTOL, has already completed its first series of flight tests in its center in Oregon (Hawkins, 2018). Besides, industry pioneers such as Rolls-Royce, Boeing, and Martin Jet Pack are making their debut into the emerging air taxi sector (Ridden, 2018).

A comprehensive summary of existing air taxi studies is exhibited in Table 1. The proposed research enables decision-makers to take several supply chain decisions, such as the number of air taxis required for efficient network operation (strategic), passenger wait time policy (tactical), and real-time routing and dispatching decisions (operational).

**Table 1:** Relevant Studies Conducted in Air Taxi Design and Operations

| Paper | Strategic | Tactical | Operational |
|---|---|---|---|
| Baik et al. (2008) | | | ✓ |
| Holden and Geol (2016) | ✓ | | ✓ |
| Duffy et al. (2017) | ✓ | | |
| Falck et al. (2018) | ✓ | | |
| Johnson et al. (2018) | ✓ | | |
| Ridden (2018) | ✓ | | |
| Sun et al. (2018) | | ✓ | ✓ |
| Hasan (2019) | | ✓ | ✓ |
| Rajendran and Zack (2019) | ✓ | | |
| Rath and Chow (2019) | ✓ | | |
| Proposed Research | ✓ | ✓ | ✓ |

**2.2 Similar Services**

**2.2.1 Ride-Sharing Taxi Services**

As a result of growing businesses in metropolitan cities, the demand for regular taxi services continues to rise. With the advancement in technology, taxi services encourage customers to participate in "ride-sharing" to attain mutual benefit. Several works provided the theoretical frameworks to offer effective ride-sharing strategies for taxi services in several metropolitan cities including Atlanta, NYC, Orlando and Singapore (Chen et al., 2010; Agatz et al., 2011 & 2012; Lin et al., 2012; Ma et al., 2013 & 2015; Ota et al., 2015; Santos and Xavier, 2015; Alonso-Mora et al., 2017; Lokhandwala and Cai, 2018; Gurumurthy and Kockelman, 2018). A majority of these studies concluded that it is useful to share taxi rides even in cities with a low "willingness to share" rate.

Many transportation network companies, such as Uber, entice the customers with ride-sharing options upfront by providing details about potential cost savings without making a significant sacrifice in time. With an upfront fare, customers know the exact price of the trip before confirming the pickup. This fare includes a base rate, a rate for estimated time and distance, which is also a function of the current demand for rides in the commuter's pickup and drop neighborhood. A booking fee and any other applicable charges are also included in the upfront fare (Posen, 2015; Witt et al., 2015).

**2.2.2 Bike Sharing Services**

Similar to air taxi services in which customers commute between vertiports and vertistops, people use bikes that are stationed in racks and ride between places with bike-sharing option. Besides the advantage of having a positive impact on the environment, bike-sharing has also been leveraged in recent years for first and last-mile commutes complementing other transportation modes such as subways, ferries, or buses (DeMaio et al., 2009). One of the critical tasks that have to be accomplished in this type of system is the development of inventory policies that strike a balance between shortage and excess bicycles at each dock. For instance, Li et al. (2015) predicted the demand for cycles at each station using a two-phased algorithm that combined the gradient boosting regression tree and clustering technique. Likewise, it is necessary to make such forecasting and inventory decisions at each station in real-time based on the geographical region incorporating seasonality.

While the above-discussed methods of transportation have a couple of similarities with the proposed air taxi services, the latter has certain unique operational constraints (as detailed in Table 2) that make the models developed in the literature not applicable for efficient air taxi operations. For instance, since air taxis cannot remain at a facility (unless the site is a vertiport) for a prolonged period of time, after each passenger drop, an immediate real-time decision has to be made on whether the air taxi has to be directed to another station or be idle at its nearest vertiport.

**Table 2:** Difference between Alternative Transportation Methods

| Diff. | Process | Regular Taxi | Subway/Buses | Bike Sharing Services | Air Taxi |
|---|---|---|---|---|---|
| 1 | *Entities involved in initiating a ride* | Driver and passenger (sometimes both these elements are given a chance to accept a ride – e.g., Uber cab allocation) | Passenger | Passenger | Passenger and logistics company (latter decides which air taxi should pick up each customer) |
| 2 | *Post customer drop-off event* | Driver decides whether to remain idle or pick up another customer | Picks up customers in that stop and travels to the next stop | Bikes are stationed at the bike stand waiting for the next customer to pick them up | Company decides on whether the air taxi has to (i) pickup customers in that facility (ii) serve customers in another site or (iii) be idle |
| 3 | *Routing decisions* | Operational | Strategic | Operational | Operational |
| 4 | *Number of stops made between customer pickup and drop-off point* | Very few (in case of ride-sharing) or no stop | Several | None | None |
| 5 | *Customer travel distance* | Short/Long | Short/Long | Mostly short | Long/Very long |
| 6 | *Potential number of locations where vehicles can halt to pickup and drop-off customers* | Enormous | Limited | Limited | Extremely limited |
| 7 | *Route taken* | Variable (based on customer destination) | Fixed | Variable | Variable |
| 8 | *Travel time uncertainty* | High | Low | High | Very Low |

The paper documents several key contributions made in the air taxi network operations domain. First, to the best of the author's knowledge, this research is the first to develop an efficient air taxi dispatching system to not only determine the effective real-time routes in a metropolitan network, but also provide insights on the number of flying vehicles required to achieve a balance between customer travel time and cost. One of the other key contributions of this research is the development of the simulation-based optimization algorithm for routing air taxis in a cyber-physical network. Studies have proven that applying optimization models for taxi dispatching is pragmatic only for small-sized problems (Wong and Bell, 2006). However, the proposed research overcomes this issue by developing a simulation-based optimization approach that can handle millions of decision variables. Third, this paper integrates a multicriteria goal programming approach into the algorithm, which will result in a solution that establishes a tradeoff between minimizing idle time and travel cost.

## 3. Data

The application of the proposed algorithm employs the data containing millions of regular yellow and green taxi records that are provided by the NYC Taxi and Limousine Commission on the government website. Each row consists of the following set of core features: (i) pickup and drop-off date and time (MM/DD/YYYY hh:mm), (ii) pickup and drop-off coordinates (latitude and longitude), (iii) trip distance (miles), and (iv) fare amount ($).

### 3.1 Sequence of Events in Air Taxi Operations

Based on the description of the air taxi network operations in Holden and Geol (2016), Hassan (2019) and Rajendran and Zack (2019), the following are the sequence of events that is expected to occur when a customer requests an air taxi ride, as depicted in Figures 2 and 3:

- *Ride request* – Customer places a ride request using the company's app in his/her smart device and enters the pickup and drop-off locations.
- *Eligibility check* – Logistics company analyzes if the individual is eligible for air taxi service and provides its client with alternate methods of commutes along with fare amounts.
- *Mode choice* – Customer decides whether to use the aviation service or not.
- If the commuter chooses to avail the air taxi service, then the transportation company has to make arrangements for the following events:
    - first leg (on-road) travel – passenger is transported by car from his/her pickup point to the source vertistop/vertiport infrastructure or walks directly to the station if it is at proximity to his/her current location.
    - soaring service – air taxi aviation travel from source to destination vertistops/vertiports.
    - last leg (on-road) travel – customer is commuted from destination site to his/her drop-off location by car or walks if the final destination is closeby.

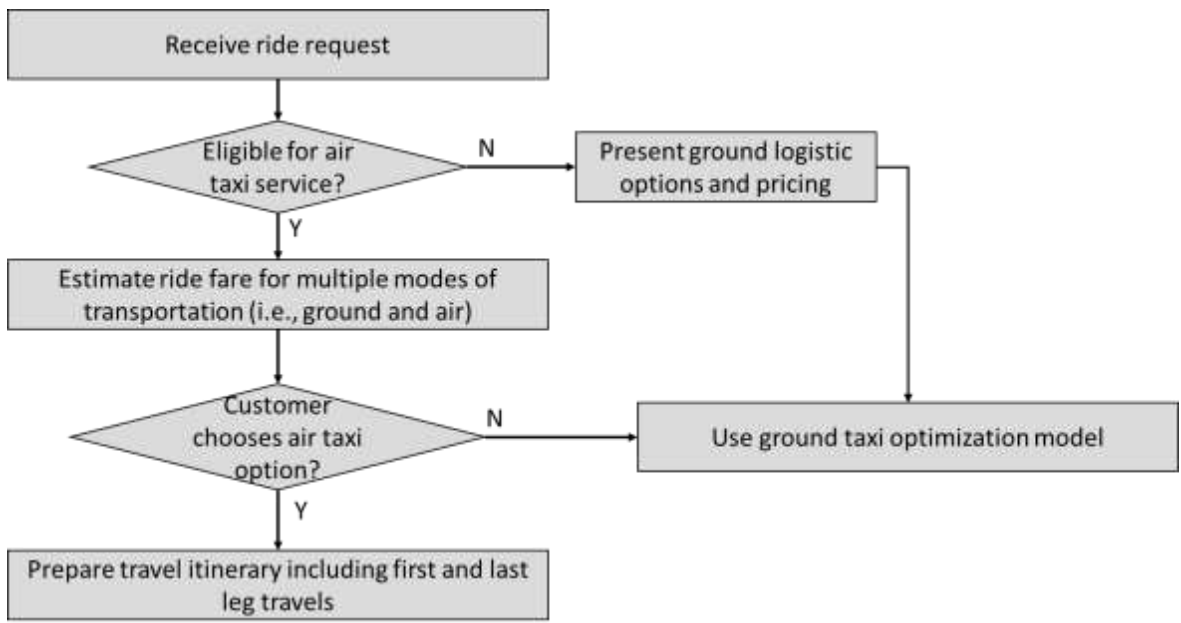

**Figure 2:** Air Taxi Booking Process

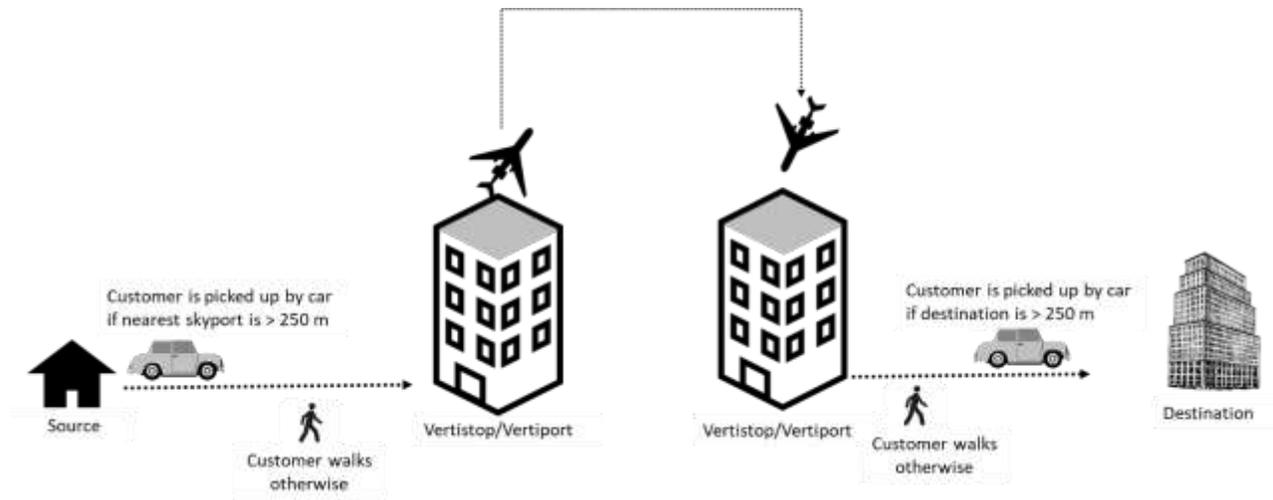

**Figure 3:** Schematic Representation of Air Taxi Network

### 3.2 Assumptions

There are two sets of assumptions considered in this paper.

Assumption Set 1 (adapted from prior studies - Holden and Geol, 2016; Duffy et al., 2017; Finger et al., 2018; Rajendran and Zack, 2019)

a) All riders are to take no more than one VTOL leg (i.e., VTOL layovers are not considered).

b) Maximum travel distance per ride is 100 miles.

c) En-route VTOL airspeed is 160 mph.

d) Takeoff and landing times are assumed to be 75 seconds each.

e) Loading and unloading time of riders take three and two minutes, respectively.

f) A rider is eligible for a VTOL route if and only if the estimated duration of the route is at least 40% faster relative to the expected duration of the ground trip (this parameter is varied in the sensitivity analysis section).

g) All requests are met on an on-demand basis (i.e., scheduling rides in advance are not considered).

h) On-road travel (i.e., first and last legs) should not be more than a mile each (this parameter is varied in the sensitivity analysis section).

i) The travel time to cover the on-road one-mile distance is 10 minutes (i.e., speed = 6 mph, estimated based on the average speed of vehicles in the NYC during rush hour, according to Furfaro et al., 2018).

Assumption Set 2 (proposed in this research)

j) Demand is stochastic and follows a Poisson arrival pattern.

k) Customers are served on the first-come, first-serve basis.

l) Vehicle maintenance and charging can be accommodated during the idle time of air taxis.

m) Travel time is in multiples of 5-minute time slots and does not take in-between values. In other words, if the actual air taxi travel time for ride $i$ is $ta_i$, then it is converted to $\left\lVert \frac{ta_i}{5} \right\rVert$. Once this categorization is done, the simulation-based optimization model is run for 288-time units, where each time unit represents a five-minute interval (i.e., total time slots/day = $\frac{1440 \text{ minutes/day}}{5 \text{ minutes/time slot}} = 288$).

### 3.3 Determining Eligible Customer Population

This section explains the methodology used by Rajendran and Zack (2019) for determining the eligible air taxi customer population.

**Notations**

Parameters

$la_i^p$  Pickup latitude of ride $i$ (in °)

$lo_i^p$  Pickup longitude of ride $i$ (in °)

$(la_i^p, lo_i^p)$  Customer pickup coordinates for ride $i$ (in °)

$la_i^d$  Drop-off latitude of ride $i$ (in °)

| | |
|---|---|
| $lo_i^d$ | Drop-off longitude of ride $i$ (in °) |
| $(la_i^d, lo_i^d)$ | Customer drop-off coordinates for ride $i$ (in °) |
| $tr_i$ | Travel time using the regular taxi service during ride $i$ (in minutes) |
| $ta_i$ | Total travel time using the air taxi service for ride $i$ (in minutes) |
| $to_i$ | Travel time of the on-road component when using the air taxi service (in minutes) |
| $tf_i$ | Flight time when customer avails the air taxi service for ride $i$ (in minutes) |
| $\Delta_i$ | Difference in time between taking the regular and air taxi services (i.e., time savings due to utilizing the air taxi option) for ride $i$ (in minutes) |
| $s$ | Estimated air taxi travel speed (in miles/minute) |

Variables

$Y_i$ $\begin{cases} 1 \text{ if ride } i \text{ is eligible for air taxi service} \\ 0 \text{ otherwise} \end{cases}$

### 3.3.1 Air Taxi Travel Time Estimation

Any air taxi ride consists of two components – the ground travel and flight travel. Thus, the estimated travel time using air taxi service $(ta_i)$ for ride $i$ is the sum of on-road travel time $(to_i)$ and flight time $(tf_i)$, as shown in Equation (1). The flight time is estimated from (a) the Haversine distance between the pickup coordinates $(la_i^p, lo_i^p)$ and the dropoff coordinates $(la_i^d, lo_i^d)$, which is represented by $\text{hav}\left((la_i^p, lo_i^p), (la_i^d, lo_i^d)\right)$, and (b) travel speed $s$, as presented in Equation (2).

$$ta_i = to_i + tf_i \tag{1}$$

$$tf_i = \frac{\text{hav}\left((la_i^p, lo_i^p), (la_i^d, lo_i^d)\right)}{s} \tag{2}$$

Where the Haversine formula to calculate the distance between two points is given using constraint (3) (adapted from Mwemezi and Huang, 2011).

$$\text{hav}\left((la_i^p, lo_i^p), (la_i^d, lo_i^d)\right) = \text{hav}(la_i^d - la_i^p) + \cos(la_i^p) \times \cos(la_i^d) \times \text{hav}(lo_i^d - lo_i^p) \tag{3}$$

### 3.3.2 Calculation of Time Savings

The time that is saved as a result of availing the air taxi service is computed as the maximum of the difference between regular and estimated air taxi travel time and 0 (Equation 4). The regular taxi travel time for each ride $i$ ($tr_i$) is provided in the NYC Taxi and Limousine Commission database.

$$\Delta_i = \max\{(tr_i - ta_i); 0\} \tag{4}$$

### 3.3.3 Eligibility Check for Availing Air Taxi Services

As mentioned in Section 3.2, each ride $i$ is eligible for the air taxi service only if the air taxi ride duration ($ta_i$) is at least 40% less than that of the ground trip ($tr_i$), as shown in constraint (5).

$$Y_i \begin{cases} 1 \text{ if } \Delta_i \geq 0.4 \times tr_i \\ 0 \text{ otherwise} \end{cases} \tag{5}$$

### 3.4 Demand Estimation

As discussed earlier, this paper adopts the pre-processed data and infrastructure location recommendations proposed by Rajendran and Zack (2019). The authors suggested a wide range of location insights under different cost, time, and demand fulfillment rate settings. With the potential sites discussed in their study, air taxi customer demands between these sites for different times of the day and days of the week are then estimated. Using this information, the proposed research aims to apply the developed simulation-based optimization algorithm for dynamically routing air taxis (as presented in Figure 4). As detailed in Section 3.2, each day is divided into 5-minute time zones (i.e., there are 288 time bins in a day), and the schematic representation of the input demand parameter is given in Table 3. Since the demand pattern varies across different times of the day and day of the week, in total, there are 453,600 (i.e., 15 sites ×15 sites × 288 time slots per day × 7 days per week) demand points.

**Table 3:** Schematic Representation of the Input Demand Data

| Day | Time Bin | Source Vertistop/Vertiport ID | Destination Vertistop/Vertiport ID | Estimated Air Taxi Demand |
|---|---|---|---|---|
| Sunday | 1 | 1 | 2 | 5 |
| Sunday | 1 | 1 | 3 | 7 |
| ... | … | … | … | … |
| Sunday | 288 | 1 | 2 | 3 |
| ... | … | … | … | … |
| Monday | 1 | 1 | 2 | 9 |
| ... | … | … | … | … |
| ... | … | … | … | … |

## 4. Methodology

In this paper, the statistical significance of the proposed model over the dynamic nearest neighbor algorithm discussed in Meesuptaweekoon and Chaovalitwongse (2014) is examined. Section 4.1 details the proposed simulation-based optimization model, and Section 4.2 describes the nearest-neighborhood algorithm. It is to be noted that the algorithms presented in this section recommend routing decisions only for air taxis and do not consider first- and last-leg on-road car scheduling.

Monte-Carlo simulation approach is used in this research to sample the arrival pattern from a specific distribution, and the average of the performance measures are reported in Table 7. Different arrival scenarios ($\omega \in \Omega$) are generated from its associated probability distribution (considered as Poisson in this study - based on the traditional queuing assumption), and each scenario corresponds to one realization of the random event, $\alpha^\omega$.

## 4.1 Simulation-Based Optimization Approach

At every time unit, the dispatch center will collect information on the estimated demand, geographical coordinates of each air taxi, idle/occupancy status, and the proposed centralized scheduling tool will route air taxis dynamically. The methodological framework is portrayed in Figure 4. Besides the objective of reducing the supply-demand mismatch, another consideration while developing the air taxi routes is to minimize the total number of air taxis that are idle at a given point of time.

The simulation-based optimization dispatching algorithm proposed in this paper (presented in Figure 5) dynamically routes each air taxi in real-time with the objective of minimizing two conflicting criteria: the number of operating air taxis and idle time. Due to the complexity involved in solving large-sized optimization models, the proposed algorithm follows a receding rolling horizon control approach. At every time unit $t$, the model decides the routing policy, taking into consideration the decisions made in the previous time slots as initial conditions. The proposed dynamic routing interface is presented in Figure 5, and an algorithmic overview of the developed simulation-based optimization model is given in Table 4.

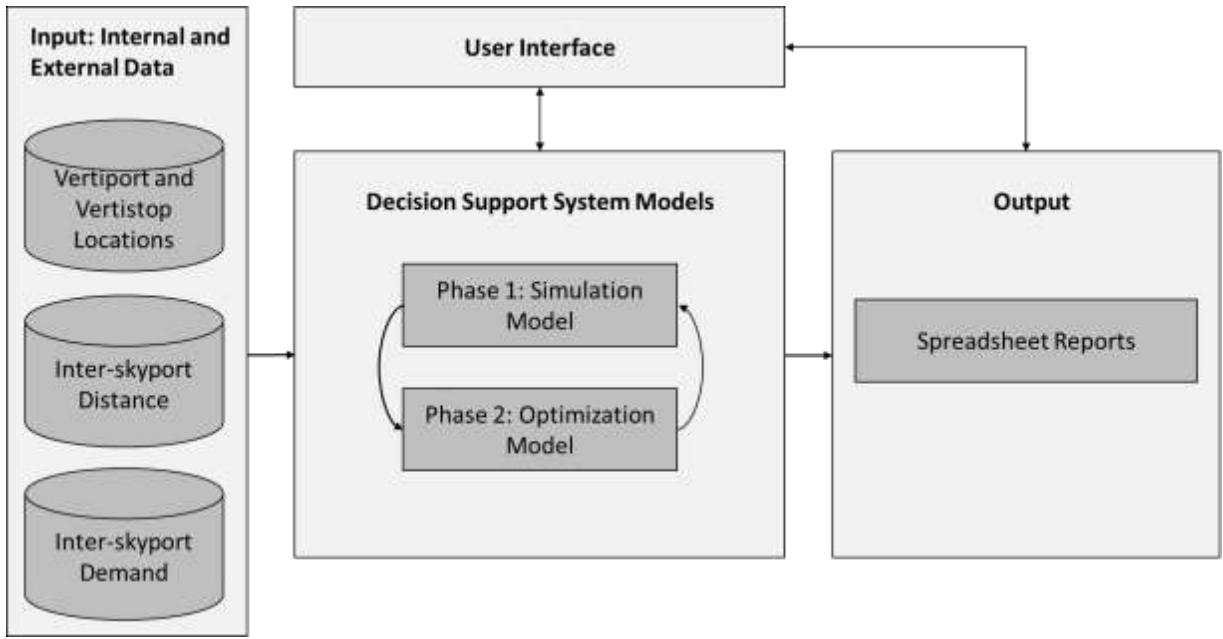

**Figure 4:** Methodological Framework

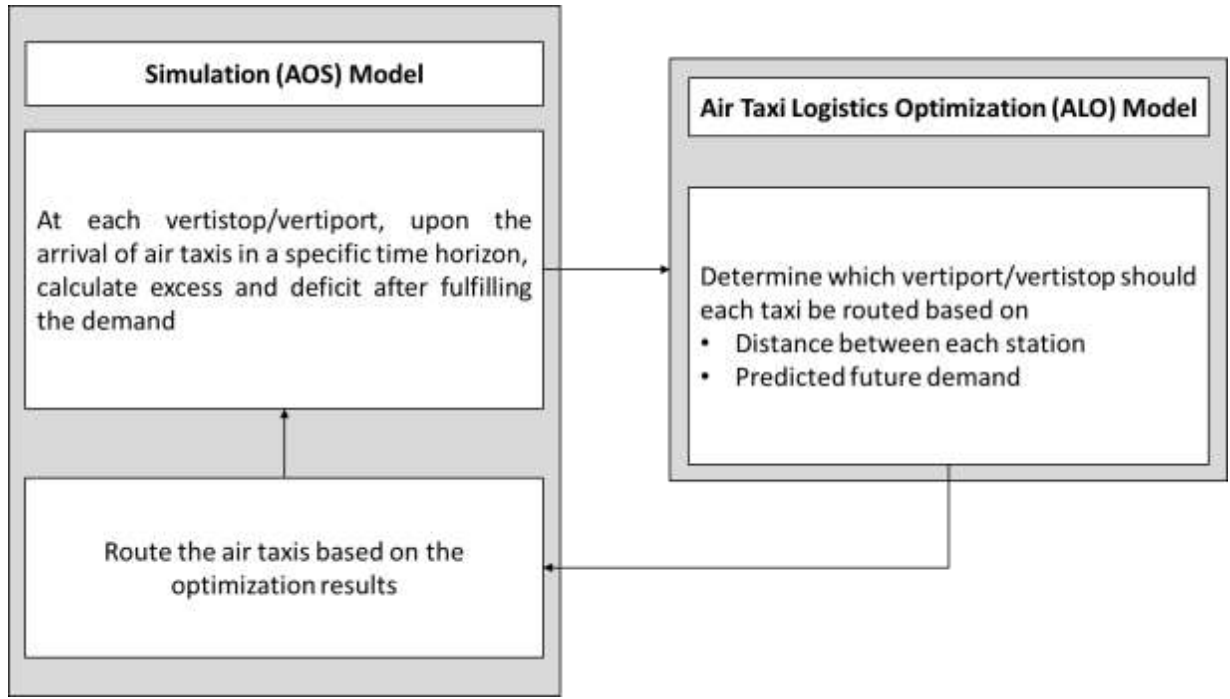

**Figure 5:** Proposed Simulation-based Optimization Dynamic Routing Interface

### 4.1.1 Notations used in Simulation-based Optimization Model

Sets and Indices:

| | |
|---|---|
| $v, v' \in V$ | Set of vertiports/vertistops |
| $t \in T$ | Time slot |
| $j \in J$ | Customer waiting time at the pickup station due to the unavailability of air taxi. For instance, if $j = 1$, then the customer has been waiting for one-time slot (i.e., 5 minutes) for the air taxi. $J$ represents the maximum allowable customer wait time. If $J = 2$, then each customer can wait at the maximum for two-time slots (= 10 minutes) at the pickup station for his/her service. |

Parameters:

| | |
|---|---|
| $d_{v,v',t}$ | Demand or the number of passengers arriving at time slot $t$ requesting rides from stations $v$ to $v'$ (note that for $v = v'$, $d_{v,v',t} = 0, \forall t$; $d_{v,v',t}$ follows a Poisson distribution) |
| $r_{v,v'}$ | Travel time between $v$ and $v'$ |
| $n(v)$ | Nearest vertiport for station $v$ (if $v$ is a vertiport by itself, then $n(v) = v$) |

Variables:

| | |
|---|---|
| $E_{v,t}$ | Excess number of air taxis available at facility $v$ at the end of time slot $t$ after fulfilling the demand |
| $U_{v,t}$ | Remaining unfulfilled demand at site $v$ in time slot $t$. Initially $U_{v,t}$ is the sum of the actual demand ($\sum_{v' \in V} d_{v,v',t}$) at $v$ during $t$, and is updated at the end of each time horizon. |
| $X_{v,v',t}$ | Number of empty air taxis assigned to travel from $v$ to $v'$ at time $t$ ($X_{v,v',t}$ is the decision variable for the ALO mathematical model). These air taxis will arrive at $v'$ at time $t + r_{v,v'}$. |

### 4.1.2 Air Taxi Operations Simulation (AOS) Model

The algorithmic overview of the developed simulation-based optimization model is given in Table 4. In each vertistop $v$, at the beginning of time slot $t$, the total number of air taxis carrying passengers from other zones arriving at $v$ are given by Step 13. Likewise, the number of empty air taxis that are arriving at vertistop $v$ at the beginning of time period $t$ is computed in Step 14, and these are air taxis that were routed to vertistop $v$ from other regions during the previous time slots (based on the optimization model discussed in Section 4.1.3). Step 15 gives the total incoming air taxis at vertistop $v$, at the beginning of every time slot $t$.

If the total incoming air taxis at a vertiport/vertistop $v$ are more than the sum of the current and backlogged demand, then the excess ($E_{v,t}$) is given in Step 17. The total number of air taxis that are dispatched from station $v$ carrying passengers to $v'$ (including passengers who have been waiting for air taxi service from the previous time slots) are given in Step 19. Since we are considering a case in which the total incoming vehicles at a vertiport/vertistop $v$ are more than the overall demand (i.e., current + leftover), all the customers waiting to avail the air taxi service are provided their rides, and hence the remaining unfulfilled demand is set to 0 in Step 24.

Whereas, if the total inventory is less than the current and backlogged demands, then passengers are served on the first-come, first-serve basis. In other words, if there are any passengers who are waiting since time slot $t - J$ (recall that $J$ is the maximum allowable time slot for which customers can wait), then they are served first, followed by those waiting since $t - j + 1$, and so on, as given in Steps 26 – 32. The backlogs from the previous time slots are updated in Step 29, and current unfulfilled demand is computed in Step 31. The model terminates if $U_{v,t-J} > 0$. That is, the number of air taxis in the planning horizon should be such that customers do not wait for more than $J$ time slots. Similar to the first case (where the inventory is less than the demand), the total number of air taxis that are dispatched from station $v$ carrying passengers to facility $v'$, are computed in given in Step 34.

After the Air Taxi Logistics Optimization (ALO) Model is ran in Step 44, the output of the mathematical model is given as an input to the AOS algorithm in Step 50. Based on the ALO model results, vehicles are dispatched either to the assigned vertiport/vertistop or sent to the vertiport where they remain idle, as shown in Steps 51 and 52.

**Table 4:** Proposed Simulation-Based Optimization Algorithm

| | For each scenario, generate demand instances, repeat steps 1 - 56 and record the output |
|---|---|
| 1 | **Initialize** number of vehicles |
| 2 | **For** $v \leftarrow 1$ to $V$ **do** |
| 3 | **Input** $n(v)$ |
| 4 | **For** $v' \leftarrow 1$ to $V$ and $v' \neq v$ **do** |
| 5 | **Input** $r_{v,v'}$ |
| 6 | **For** $t \leftarrow 1$ to $T$ **do** |
| 7 | **Input** $d_{v,v',t}$ and $U_{v,t}$ |
| 8 | **End for** |
| 9 | **End for** |
| 10 | **End for** |
| 11 | **For** $t \leftarrow 1$ to $T$ **do** |
| 12 | **For** $v \leftarrow 1$ to $V$ **do** |
| 13 | **Compute** total air taxis entering vertiport/vertistop $v$ at time slot $t$ carrying passengers from all other stations, $I_{v,t}^1 = \sum_{\substack{v' \in V \\ v' \neq v}} \left( d''_{v',v,t-r_{v,v'}} \right)$ |
| 14 | **Compute** total empty air taxis entering site $v$ at time slot $t$ from all other facilities, $I_{v,t}^2 = \sum_{\substack{v' \in V \\ v' \neq v}} X_{v',v,t-r_{v,v'}}$ |
| 15 | **Compute** the total units of incoming air taxis during time slot $t$ at vertiport/vertistop $v$, $I'_{v,t} = I_{v,t}^1 + I_{v,t}^2$ |
| 16 | **If** the total incoming air taxis are more than the sum of demand at the current time slot and backlogged unfulfilled demand (i.e., if $I'_{v,t} > (\sum_{\substack{v' \in V \\ v' \neq v}} d_{v,v',t} + \sum_{j \in J} U_{v,t-j})$, **then** |

| 17 | **Compute** excess air taxi at time slot $t$, $E_{v,t} = I'_{v,t} - \sum_{\substack{v' \in V \\ v' \neq v}} d_{v,v',t} - \sum_{j \in J} U_{v,t-j}$ |
|---|---|
| 18 | **For** $v' \leftarrow 1$ to $V$ and $v' \neq v$ **do** |
| 19 | **Compute** the total air taxis that are dispatched from site $v$ carrying passengers to $v'$ at time slot $t$, $d''_{v,v',t} = d_{v,v',t} + \sum_{j \in J} d'_{v,v',t-j}$ |
| 20 | **End for** |
| 21 | **For** $i \leftarrow 1$ to $T - t$ **do** |
| 22 | **Update** unfulfilled demand at $t + i$, $U_{v,t+i} \leftarrow U_{v,t+i} - \sum_{v' \in V \mid r_{v,v'} = i} d''_{v,v',t}$ |
| 23 | **End for** |
| 24 | **Update** backlog demands to 0 (i.e., $\sum_{j \in J} U_{v,t-j} = 0$; $U_{v,t} = 0$ and $\sum_{\substack{v' \in V \\ v' \neq v}} \sum_{j \in J} d'_{v,v',t-j} = 0$) |
| 25 | **Else** |
| 26 | **Assign** air taxis on a first-come, first-serve basis to the customers and set the excess air taxi at station $v$ during time slot $t$ to 0 (i.e., $E_{v,t} = 0$) |
| 27 | **For** $v' \leftarrow 1$ to $V$ and $v' \neq v$ **do** |
| 28 | **For** $j \leftarrow 1$ to $J$ **do** |
| 29 | **Update** backlog of previous time slots, $d'_{v,v',t-j} \leftarrow d'_{v,v',t-j} - I''_{v,v',j,t}$ (where $I''_{v,v',j,t}$ is defined as the proportion of the total inventory that is used to fulfill the customer demand from $v$ to $v'$ that occurred during time slot $t - j$) |
| 30 | **End for** |
| 31 | **Compute** backlog of current time slot $t$, $d'_{v,v',t} \leftarrow d_{v,v',t} - I''_{v,v',0,t}$ (where $I''_{v,v',0,t}$ is the proportion of the total inventory that is used to fulfill the customer demand from $v$ to $v'$ that occurred during time slot $t$. Since the demand is fulfilled on a FIFO basis, demand at the current time slot $t$ will be given the least priority. Note that $\sum_{j \in J} I''_{v,v',j,t} + I''_{v,v',0,t} = I'_{v,t}$) |
| 32 | **End for** |
| 33 | **For** $v' \leftarrow 1$ to $V$ and $v' \neq v$ **do** |

| 34 | **Compute** the total air taxis that have dispatched from $v$ carrying passengers to $v'$ at time slot $t$, $d''_{v,v',t} = I''_{v,v',0,t} + \sum_{j \in J} I''_{v,v',j,t}$ |
|---|---|
| 35 | **End for** |
| 36 | **For** $j \leftarrow 0$ to $J$ **do** |
| 37 | **Compute** the total backlog at $v$ at each time slot, $U_{v,t-j} = \sum_{\substack{v' \in V \\ v' \neq v}} d_{v,v',t-j}$ |
| 38 | **End for** |
| 39 | **If** $U_{v,t-J} > 0$, **then** |
| 40 | **STOP** |
| 41 | **End if** |
| 42 | **End if** |
| 43 | **End for** |
| 44 | **Run** Air Taxi Logistics Optimization (ALO) Model |
| 45 | **For** $v \leftarrow 1$ to $V$ **do** |
| 46 | **For** $v' \leftarrow 1$ to $V$ and $v' \neq v$ **do** |
| 47 | **Input** $X_{v,v',t}$ (this is the output obtained from running the ALO Model) |
| 48 | **End for** |
| 49 | **End for** |
| 50 | **For** $v \leftarrow 1$ to $V$ **do** |
| 51 | **Dispatch** air taxis based on the ALO model results, and update the excess at each vertiport/vertistop $v$, $E_{v,t} \leftarrow E_{v,t} - \sum_{\substack{v' \in V \\ v' \neq v}} X_{v,v',t}$ |
| 52 | **Direct** the remaining air taxis to the nearest vertiport, $n(v)$, $E_{n(v),t+r_{v,n(v)}} \leftarrow E_{n(v),t+r_{v,n(v)}} + E_{v,t}$ (they, however, reach the nearest port only after $t + r_{v,n(v)}$ time units) |
| 53 | **For** $i \leftarrow 1$ to $T - t$ **do** |
| 54 | **Update** the remaining unfulfilled demand $U_{v,t+i} \leftarrow U_{v,t+i} - \sum_{v' \in V \mid r_{v,v'}=i} X_{v',v,t}$ |
| 55 | **End for** |
| 56 | **End for** |

### 4.1.3 Air Taxi Logistics Optimization (ALO) Model

With the estimated values of the variables $E_{v,t}$ and $U_{v,t}$ at each vertistop from the AOS model, the mathematical model optimizes air taxi routing and dispatching for a specific time period $t$. Naturally, the intent is to reduce the idle time and the shortage of the air taxis, and hence, the model looks ahead for the futuristic estimated demand and routes the air taxi to the zones where there is expected to be a vehicle deficit in the future. As discussed earlier, for every time unit under consideration, the model dynamically addresses the following:

i. Whether each air taxi should remain operational or become idle (e.g., during lean periods, certain air taxis will become idle)?

ii. If the air taxi is functional, to which vertistop/vertiport should it be routed to, taking into consideration the future customer demand?

The overall goal of this research is to estimate the optimal number of air taxis required for the efficient functioning of the network. Hence, at any given time slot $t$, the objectives of the ALO model are (i) to reduce the number of idle air taxis (Equation 6), and (ii) to minimize the total distance traveled by air taxis at an inactive state in the cyber-physical system (Equation 7). Clearly, these objectives are conflicting, and the preemptive goal programming technique is used to solve the multi-criteria problem.

#### 4.1.3.1 Preemptive Goal Programming

In the goal programming method, the objective functions are assigned as goals. Each goal possesses a pre-specified tolerance (i.e., in this case, a value for the number of idle air taxis and the total distance traveled by all air taxis). These figures are only targets and can be fulfilled with some allowable deviations, and the aim of the goal programming model is to minimize these deviations from their respective targets. Hence, this approach attempts to deliver a solution that is closest to the targets, and the output is referred to in the literature as the "most preferred solution" (Srinivas et al., 2016; Rajendran et al., 2019).

In the preemptive goal programming method, the goals are categorized as high and low priorities, and the critical goal is considered absolutely more important than the smaller priority goal. The preemptive goal programming model consists of two stages; in stage-1, the model is solved by only considering the highly critical goal (which in the current problem is to reduce the total number of idle air taxis). While in stage-2, the model is solved, taking into account the low-priority goal (i.e., the total distance traveled by air taxis at an idle state), setting the alternate optima from stage-1 as the solution space for stage-2 (Jensen and Bard, 2003).

Equation (8) guarantees that the total units dispatched from each vertiport/vertistop must be less than the availability. Equation (9) ensures that the total units arriving at vertiport/vertistop $v$ in time $t + i$ are less than the required number of air taxis at that time. Equation (10) takes care of the positive integer constraint. It is important to note that <u>$t$ is a fixed value in the ALO model and is incremented only in the AOS algorithm</u>. In other words, for each value of $t$, the ALO model is solved, and $t$ incremented only after the entire simulation-optimization model is run for a time slot.

$$\text{Min } IT_t = \sum_{v \in V} \left( E_{v,t} - \sum_{\substack{v' \in V \\ v' \neq v}} X_{v,v',t} \right) \tag{6}$$

$$\text{Min } TD_t = \sum_{v \in V} \sum_{\substack{v' \in V \\ v' \neq v}} r_{v,v'} \times X_{v,v',t} \tag{7}$$

$$\sum_{\substack{v' \in V \\ v' \neq v}} X_{v,v',t} \leq E_{v,t} \qquad \forall v \tag{8}$$

$$\sum_{v' \in V | r_{v,v'} = i} X_{v',v,t} \leq U_{v,t+i} \qquad \forall v, i \leq T - t \tag{9}$$

$$X_{v,v',t} \in Z^+ \qquad\qquad \forall v, v' \qquad (10)$$

## 4.2 Nearest Neighbor Algorithm

In the nearest neighbor algorithm, after dropping off the customers, excess vehicles at each vertiport/vertistop are dispatched to the nearest station in which there is expected to exist an air taxi deficit in the next time period. In other words, in contrast to the proposed AOS-ALO approach, this algorithm bases its decision looking only at the following time slot and not at any other future time zones. After assigning, the remaining idle vehicles are sent to the nearest vertiport and remain unused. A brief overview of the algorithm is presented below.

| | |
|---|---|
| 1 | **Initialize** parameters for each scenario |
| 2 | **Compute** excess and surplus air taxis at each station at the end of time $t$ |
| 3 | **Assign** air taxis to the nearest station where vehicles are required for the next time horizon (nearest neighbor, in this case, is defined as the station that can be reached within one-time slot) |
| 4 | **Reroute** unassigned air taxis to the nearest vertiport |
| 5 | **Update** time unit |
| 6 | **Repeat** Steps 2 to 5 |

## 5. Results

This section details the findings obtained using the simulation-based optimization model and the algorithm from the literature. Sensitivity analysis is conducted to explore the impact of the percentage of time savings, willingness to fly rate, on-road travel limit, demand fulfillment rate, maximum customer wait time, and arrival distribution on the performance measures. The algorithm was coded in Python®, and the ALO model was solved using IBM CPLEX® 12.8 optimizer.

**5.1 Data Cleaning and Demand Estimation**

For a span of two years, Rajendran and Zack (2019) estimated that about 4.7 million customers (i.e., nearly 6500 commuters/day) would be eligible for availing the air taxi service at NYC, confined to the settings discussed in Section 3.2. Whereas, six million commuters are expected to use this service if the demand fulfillment rate is set at 100%. The estimated average daily demand for air taxi is portrayed in Figure 6. We can notice that the demand is lowest during the weekends, which is expected because of the reduction in passengers traveling for work, while it is the highest on Thursdays and Fridays.

Figure 7 shows the average potential air taxi demand during the course of a day. The data reveals that the demand for air taxis is the lowest between 12 AM and 6 AM. Unexpectedly, it touches the peak during 3:00 PM – 5:30 PM, in contrast, the calls for regular green and yellow taxis at NYC is at its highest after 5:30 PM (Sebastian, 2018). From Figure 8 that exhibits the frequency of the distance traveled by the potential air taxi customers on a daily basis, it can be inferred that most of the customers travel between 15 and 20 miles on average.

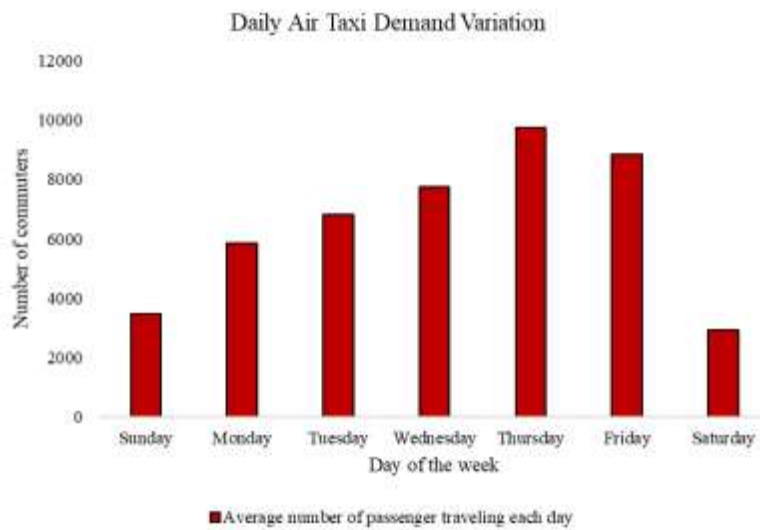

**Figure 6:** Estimated Daily Air Taxi Demand Variation

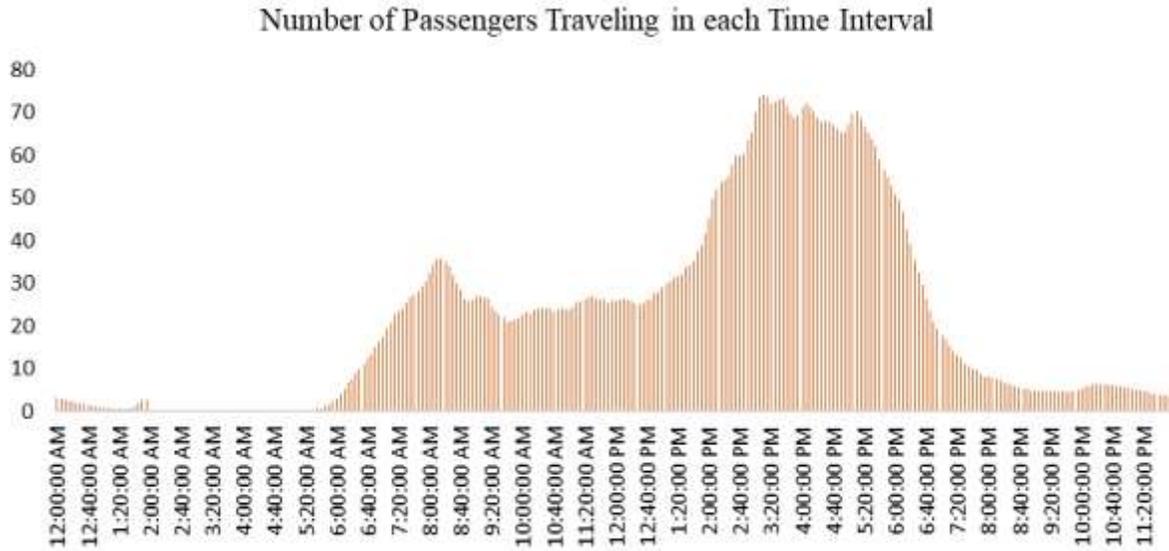

**Figure 7:** Number of Estimated Air Taxi Passengers Traveling in each Time Interval

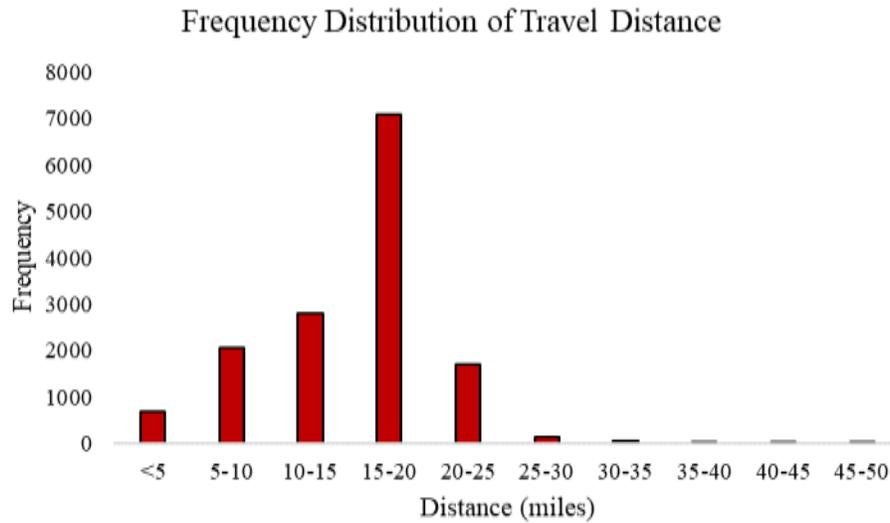

**Figure 8:** Histogram of Estimated Air Taxi Passenger Travel Distance

**5.2 Vertistop and Vertiport Locations in New York City**

Based on the assumptions listed in Section 3.2, the study by Rajendran and Zack (2019) recommended establishing 2 vertiports and 16 vertistops across different regions in NYC. The locations of the vertiports were at South Central Park and JFK International Airport (depicted by green triangles in Figure 9), while it was suggested to launch vertistops at the several zones including landmarks, such as Newark Liberty International and LaGuardia Airports, World Trade Center, and North Central Park (vertistops are shown by red triangles). Among the findings, three infrastructure sites were outside NYC with a very insignificant coverage rate, and hence, this research does not consider those sites with insufficient demand. The 15 sites under consideration are detailed in Table 5.

In addition to considering the above-specified details as the first set of inputs to the proposed algorithm, the other two required data are the following (as portrayed in Figure 9): (a) the distance between the 15 infrastructure sites given in Table 5, which will be a matrix with a dimension of $15 \times 15$; and (b) demand between vertiports/vertistops for every 5-minute interval (i.e., 288 time slots in a day) for different days of the week in a format shown in Table 3.

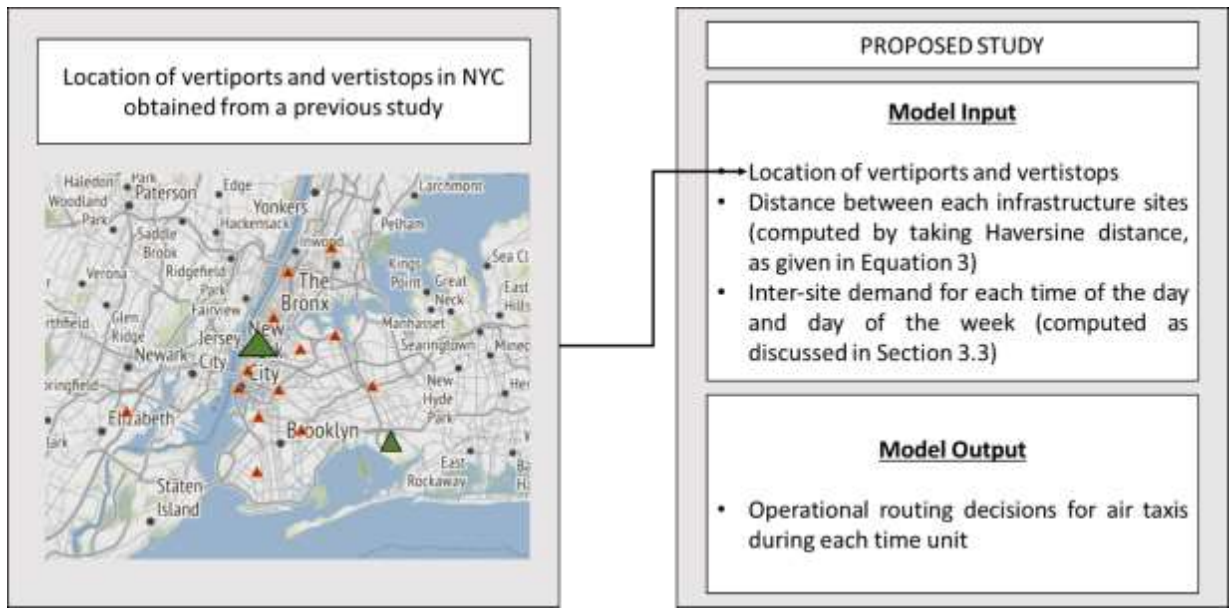

**Figure 9:** Overview of the Proposed Study

**Table 5:** Location of the Recommended Air Taxi Sites

| Site ID | Type of Facility | Location |
|---|---|---|
| 1 | Vertiport | South Central Park, Manhattan |
| 2 | Vertistop | Douglass Street, Brooklyn |
| 3 | Vertistop | South Congress Ave, Bronx |
| 4 | Vertistop | 43$^{rd}$ Street, Long Island City |
| 5 | Vertiport | JFK International Airport, Queens |
| 6 | Vertistop | 84$^{th}$ Avenue, Jamaica |
| 7 | Vertistop | 62$^{nd}$ Street, Brooklyn |
| 8 | Vertistop | Grafton Street, Brooklyn |
| 9 | Vertistop | 97$^{th}$ Street Transverse, New York |
| 10 | Vertistop | Newark International Airport |
| 11 | Vertistop | One World Trade Center, New York |
| 12 | Vertistop | Williamsburg, Brooklyn |
| 13 | Vertistop | Audubon Avenue, New York |
| 14 | Vertistop | LaGuardia International Airport |
| 15 | Vertistop | Washington Square North, New York |

**5.3 Base Case Results**

In the preliminary study, considering the assumptions discussed in Section 3.2, the number of flying taxis required for the network operation is found to be 84, with an average utilization rate of nearly 66% (as given in Table 7). In other words, for almost 34% of the time, the air taxis remain idle, which can perhaps be used for vehicle charging and maintenance (such as safety testing and cleaning). On average, nearly 6500 commuters can leverage this soaring service, and each air taxi serves about 77 customers per day in NYC. Almost 10% of the total demand is between vertiports located near South Central Park (site #1) and the site in LaGuardia International Airport (site #14). While facilities in Manhattan cover a major proportion of the air taxi market demand, about 92% of the time, there is no customer demand at Bronx (at site #4). The demand density is less even in Staten Island and the southern zones of Brooklyn.

The inter-site demand is the least during the late nights and early mornings. In fact, between 11 PM and 2 AM, the maximum number of customers requesting rides between any two sites is only one (mainly observed in Manhattan), with no customer demand for 86% of this three-hour time interval. In a 5-minute time slot, the number of customers requesting air taxi rides between two sites is as high as seven passengers, mainly traveling from 9 - 10 AM and 5 - 6 PM. While more people are going from LaGuardia Airport to Manhattan in the morning, it is observed that there is significant travel made from JFK Airport to Central Manhattan during the evening hours (which is unforeseen as one would anticipate more commuters entering the city during the morning hours).

## 5.4 Sensitivity Analysis

In this section, the percentage of time savings, willingness to fly rate, on-road travel limit, demand fulfillment rate, maximum customer wait time, and arrival distribution are varied, and their impacts on the optimal number of air taxis, utilization rate and the number of customers served are explored. The settings are discussed in Table 6. The passenger-incurred costs using the regular taxi ($RT$) and air taxi ($AT$) are then compared, and the percentage difference $\left(\left(\frac{cost_{AT} - cost_{RT}}{cost_{AT}}\right) \times 100\%\right)$ is reported in Table 7. While the passenger-incurred regular taxi cost for each ride is provided by the NYC Taxi and Limousine Commission along with the other information, the air taxi cost per passenger is calculated based on the near-term air-taxi operational cost of $5.73 per passenger mile, as reported in several articles (e.g., Holden and Geol, 2016; Binder et al., 2018; Garrow et al., 2019), and is given in Equation (11).

$$\text{Air taxi cost per passenger} = \frac{\$5.73}{\text{passenger mile}} \times \text{average dist. traveled per passenger (in miles)} \qquad (11)$$

**Table 6:** Sensitivity Analysis Settings

| Setting | Time Savings (TS) | Passenger's "willingness to fly" Rate (PR) | On-Road Travel Limit (RL) | Percentage of Customer Demand Fulfilled (CF) | Maximum Customer Wait Time (WT) | Arrival Distribution of Passengers (AD) |
|---|---|---|---|---|---|---|
| **Time Savings** | | | | | | |
| TS-1 | 30% | 100% | 1 mile | 70% | 10 minutes | Poisson |
| TS-2 (Base Case) | 40% | 100% | 1 mile | 70% | 10 minutes | Poisson |
| TS-3 | 50% | 100% | 1 mile | 70% | 10 minutes | Poisson |
| TS-4 | 60% | 100% | 1 mile | 70% | 10 minutes | Poisson |
| **Passengers Willingness to Fly Rate** | | | | | | |
| PR-1 (Base Case) | 40% | 100% | 1 mile | 70% | 10 minutes | Poisson |
| PR-2 | 40% | 90% | 1 mile | 70% | 10 minutes | Poisson |
| PR-3 | 40% | 80% | 1 mile | 70% | 10 minutes | Poisson |
| PR-4 | 40% | 70% | 1 mile | 70% | 10 minutes | Poisson |
| **On-Road Travel Limit** | | | | | | |
| RL-1 | 40% | 100% | 0.5 mile | 70% | 10 minutes | Poisson |
| RL-2 (Base Case) | 40% | 100% | 1 mile | 70% | 10 minutes | Poisson |
| RL-3 | 40% | 100% | 1.5 mile | 70% | 10 minutes | Poisson |
| **Customer Demand Fulfillment Rate** | | | | | | |
| CF-1 | 40% | 100% | 1 mile | 60% | 10 minutes | Poisson |
| CF-2 (Base Case) | 40% | 100% | 1 mile | 70% | 10 minutes | Poisson |
| CF-3 | 40% | 100% | 1 mile | 80% | 10 minutes | Poisson |
| CF-4 | 40% | 100% | 1 mile | 90% | 10 minutes | Poisson |
| **Maximum Customer Wait Time** | | | | | | |
| WT-1 | 40% | 100% | 1 mile | 70% | 5 minutes | Poisson |
| WT-2 (Base Case) | 40% | 100% | 1 mile | 70% | 10 minutes | Poisson |
| WT-3 | 40% | 100% | 1 mile | 70% | 15 minutes | Poisson |
| **Arrival Distribution** | | | | | | |
| AD-1 (Base Case) | 40% | 100% | 1 mile | 70% | 10 minutes | Poisson |
| AD-2 | 40% | 100% | 1 mile | 70% | 10 minutes | Uniform |
| AD-3 | 40% | 100% | 1 mile | 70% | 10 minutes | Normal |

### 5.4.1 Impact of the Percentage of Time Savings (TS)

The base case operates with a premise that a customer is eligible for an air taxi ride only if the time savings is more than 40% (compared to regular taxi ride). To understand the impact of this parameter, the time savings percentage is varied between 30% and 60%, in increments of 10%, as shown in Table 6. Results in Table 7 confirm that time savings negatively impacts the required number of air taxis. This can be justified because, with the increase in the percentage of time savings parameter, the number of eligible riders decreases, which in turn leads to a reduction of the required number of air taxis. Contrary to one's expectation that the utilization rate would remain almost the same across all the TS settings (as the decrease in air taxis compensate for the decline in the eligible pool of commuters), there exists a steady decline in the efficiency metric. It is also interesting to note that the number of customers per air taxi ride declines almost linearly with a linear rise in time savings.

Under the TS-1 setting, the maximum demand is observed in LaGuardia Airport and South Central Park. Also, the latter site has the most number of incoming passengers from different regions, with the majority traveling from LaGuardia Airport. In TS-3, 6.9% of the demand is occurring from facilities at LaGuardia Airport to the intersection of $8^{th}$ Avenue & $50^{th}$ Avenue (i.e., Midtown Manhattan), whereas only 3% of the demand is taking place vice versa. In TS-4, the coverage rate at each facility decreases, with 52% of the demand is outgoing from LaGuardia Airport, and the maximum customers traveling from one station to another in a 5-minute interval is only four (compared to seven in the base case).

### 5.4.2 Impact of Passenger's "willingness to fly" Rate

The perception or passenger's "willingness to fly" rate (PR) is used to take into consideration the proportion of customers who might not be interested in availing the air taxi service (for certain reasons such as not willing to go over the hassle of using multi-modal transport, acrophobia and reluctant to pay extra) even though they are eligible. The PR was assumed to be 100% (PR-1) in the base case and is changed to 90% (PR-2), 80% (PR-3), and 70% (PR-4), as given in Table 6. We can see that with the steady decrease in passenger's "willingness to fly" rate, the number of air taxis required to maintain the desired service level as well as the utilization decrease linearly. It is also essential to highlight the fact that the number of customers served per air taxi remains almost the same as in the base case.

Nearly 16% of the total demand occurs between Midtown Manhattan and JFK International Airport. In PR-2, one of the vertistops located near Jamaica Hills remains idle without customers for 44% of the regular working hours (i.e., from 9:00 AM – 6:00 PM). Other less utilized sites include sites in Brooklyn, Allerton Ballfields, and Washington Heights in Upper Manhattan. In contrast, facilities near Bryant Park, JFK International Airport, Washington Square Park and City Hall Park are used more than 50% of the time.

### 5.4.3 Impact of On-road Travel Limit

The restriction discussed in Section 3.2 that each commuter can travel at the maximum of a mile during the first and last leg on-road commutes is altered from 0.5 (RL-1) to 1.5 miles (RL-3), as described in Table 6. One would foresee that due to the decrease in the number of stations, the on-road travel time of customers increases, which might, in turn, lead to a reduction in the eligible customer pool (owing to the 40% time-savings constraint). Nonetheless, the number of customers served and the required number of air taxis is almost the same across the different RL settings. This is indicative of the fact that even if a significant portion of the travel is on the road, customers still stand to gain using air taxi services. However, the average utilization of the vehicles almost doubles from settings RL-1 to RL-3. An explanation for this finding is that RL-1 constraints the on-road travel distance to 0.5 miles, and thus attributing to the average travel time incurred by customers to about 22 minutes, on the contrary, it is nearly 31 and 41 minutes respectively for RL-2 (base case) and RL-3. Consequently, for almost the same number of vehicles and customers, the utilization increases.

Stations at JFK International Airport, Midtown East and the Museum of Modern Art serve a significant portion of the passengers under RL-1, whereas sites in the East of New York have a coverage rate of less than 2% in RL-3. RL-3 has fewer facilities and more than 40% of the total passengers commute between sites at the intersection of 45$^{th}$ Street & 6$^{th}$ Avenue and JFK International Airport, while less than 1% of the total rides are made to and from stations in the South of Brooklyn and Jerome Park Library in the Bronx.

### 5.4.4 Impact of Demand Fulfillment Rate

The preliminary results are obtained under the condition that the total distance traveled on road by 70% of air taxi passengers should not more than one mile individually for first- and last-leg travel. This service rate is varied from 60% to 100%, in increments of 10%, as given in Table 6. As the demand fulfillment rate rises by 10%, the potential customer population expands. A similar pattern is observed with the required number of air taxis, which increases on average by about 17%. As a result, almost a linear decline in utilization, as well as customers per air taxi per day, are observed. Interestingly, with the rise in demand fulfillment rate, more demand is seen in the southern districts of Brooklyn and the Bronx.

**Table 7:** Model Results for Different Settings

| Setting | # of Vertiports/Verstops[†] | # of air taxis | Average # of customers/day/ air taxi | Avg. Utilization (%) | Passenger-incurred cost ($/ride/commuter) | | |
|---|---|---|---|---|---|---|---|
| | | | | | Air Taxi | Regular Taxi[†] | Percentage deviation (%) |
| **Time Savings** | | | | | | | |
| TS-1 | 16 | 126 | 94 | 80.83 | 79.71 | 31.46 | 60.53 |
| TS-2 (Base) | 15 | 84 | 77 | 65.85 | 86.87 | 33.49 | 61.45 |
| TS-3 | 15 | 44 | 62 | 53.42 | 93.00 | 34.77 | 62.61 |
| TS-4 | 14 | 15 | 47 | 40.49 | 97.62 | 36.04 | 63.08 |
| **Passengers Willingness to Fly** | | | | | | | |
| PR-1 (Base) | 15 | 84 | 77 | 65.85 | 86.87 | 33.49 | 61.45 |
| PR-2 | 18 | 75 | 78 | 66.46 | 86.87 | 33.31 | 61.66 |
| PR-3 | 18 | 68 | 76 | 64.94 | 86.76 | 33.26 | 61.66 |
| PR-4 | 17 | 59 | 75 | 63.92 | 86.87 | 33.29 | 61.68 |
| **On-road Travel Limit** | | | | | | | |
| RL-1 | 54 | 91 | 69 | 42.01 | 87.91 | 33.43 | 61.97 |
| RL-2 (Base) | 15 | 84 | 77 | 65.85 | 86.87 | 33.49 | 61.45 |
| RL-3 | 10 | 76 | 85 | 72.48 | 87.68 | 33.41 | 61.90 |
| **Customer Demand Fulfillment Rate** | | | | | | | |
| CF-1 | 12 | 70 | 81 | 69.46 | 88.61 | 33.56 | 62.12 |
| CF-2 (Base) | 15 | 84 | 77 | 65.85 | 86.87 | 33.49 | 61.45 |
| CF-3 | 28 | 104 | 69 | 59.57 | 85.95 | 33.25 | 61.31 |
| CF-4 | 72 | 123 | 66 | 55.88 | 84.68 | 33.22 | 60.77 |
| **Maximum Customer Wait Time** | | | | | | | |
| WT-1 | 15 | 103 | 63 | 53.86 | 86.87 | 33.49 | 61.45 |
| WT-2 (Base) | 15 | 84 | 77 | 65.85 | 86.87 | 33.49 | 61.45 |
| WT-3 | 15 | 63 | 103 | 88.20 | 86.87 | 33.49 | 61.45 |
| **Arrival Distribution** | | | | | | | |
| AD-1 (Base) | 15 | 84 | 77 | 65.85 | 86.87 | 33.49 | 61.45 |
| AD-2 | 15 | 83 | 76 | 64.58 | 86.87 | 33.49 | 61.45 |
| AD-3 | 15 | 84 | 78 | 66.02 | 86.87 | 33.49 | 61.45 |

[†]reported by Rajendran and Zack (2019)

### 5.4.5 Impact of Maximum Customer Wait Time

The base case is constricted to the condition that all customers can wait for a maximum of 10 minutes for air taxi pickup at the station. This wait time parameter is varied from five minutes (WT-1) to fifteen minutes (WT-3). As anticipated, when customers can wait for more time units (i.e., while transitioning from settings WT-1 to WT-3), fewer air taxis are required to serve the same number of customers. This, in turn, leads to an increase in vehicle utilization. The passenger-incurred cost, however, remains the same since it is estimated as in terms of "$/ride/commuter".

### 5.4.6 Impact of Arrival Distribution

Consistent with the traditional queuing models, the base case assumes that the customer arrival follows a Poisson distribution. In this section, the influence of the arrival distribution on the output measures is investigated. Setting AD-1 is the base case (i.e., arrivals following Poisson distribution), while AD-2 and AD-3 assume the arrival to follow Uniform and Normal distributions respectively.

The comparisons reveal that results under the Uniform distribution arrival pattern achieve a statistically insignificant reduction in the utilization and the required number of air taxis. The findings align with the fact that Uniform distribution bounds the arrival values generated using its parameters, unlike Normal and Poisson distributions that generate a more extensive range of values.

### 5.5 Comparison of the Proposed Method with Nearest Neighborhood Algorithm

This section summarizes the findings of the developed method and the nearest neighborhood algorithm. Table 8 presents the performance measure values obtained under the two techniques. The comparison of the results disclosed that the proposed model achieves a statistically significant improvement in the number of air taxis and average utilization performs compared to the nearest neighbor algorithm. Whereas, the latter heuristic takes significantly lower time to yield the outputs, and hence making it computationally less complex.

**Table 8: Comparison of Proposed Simulation-Based Optimization Algorithm with Nearest Neighbor Heuristic**

| Setting | Proposed Simulation-Based Optimization Approach | | | Nearest Neighbor Algorithm | | |
|---|---|---|---|---|---|---|
| | # of air taxis* | Avg. Utilization%* | Computational Time (in secs) | # of air taxis | Avg. Utilization% | Computational Time (in secs)* |
| **Time Savings** | | | | | | |
| TS-1 | 126 | 80.83 | 5357 | 189 | 55.96 | 2252 |
| TS-2 (Base) | 84 | 65.85 | 4090 | 130 | 52.85 | 1603 |
| TS-3 | 44 | 53.42 | 3790 | 67 | 34.51 | 1547 |
| TS-4 | 15 | 40.49 | 2178 | 19 | 31.02 | 732 |
| **Passengers Willingness to Fly** | | | | | | |
| PR-1 (Base) | 84 | 65.85 | 4090 | 130 | 52.85 | 1603 |
| PR-2 | 75 | 66.46 | 3870 | 109 | 46.09 | 1681 |
| PR-3 | 68 | 64.94 | 3596 | 105 | 43.67 | 1647 |
| PR-4 | 59 | 63.92 | 3619 | 88 | 43.03 | 1609 |
| **On-road Travel Limit** | | | | | | |
| RL-1 | 85 | 45.67 | 3623 | 132 | 35.98 | 1507 |
| RL-2 (Base) | 84 | 65.85 | 4090 | 130 | 52.85 | 1603 |
| RL-3 | 82 | 88.92 | 4476 | 126 | 65.37 | 1762 |
| **Customer Demand Fulfillment Rate** | | | | | | |
| CF-1 | 70 | 69.46 | 4025 | 112 | 48.29 | 2712 |
| CF-2 (Base) | 84 | 65.85 | 4090 | 130 | 52.85 | 1603 |
| CF-3 | 104 | 59.57 | 4545 | 134 | 48.12 | 1428 |
| CF-4 | 123 | 55.88 | 4567 | 150 | 47.99 | 1195 |
| **Maximum Customer Wait Time** | | | | | | |
| WT-1 | 103 | 53.86 | 3412 | 153 | 40.92 | 1510 |
| WT-2 (Base) | 84 | 65.85 | 4090 | 130 | 52.85 | 1603 |
| WT-3 | 63 | 88.20 | 5140 | 113 | 71.45 | 1812 |
| **Arrival Distribution** | | | | | | |
| AD-1 (Base) | 84 | 65.85 | 4090 | 130 | 52.85 | 1603 |
| AD-2 | 83 | 65.88 | 4101 | 132 | 51.09 | 1599 |
| AD-3 | 84 | 66.02 | 4107 | 132 | 51.11 | 1681 |

*Indicates that the method performs significantly better than the other for that particular measure at α=0.05

## 5.6 Managerial Insights

Figure 10 gives the value path graph, which enables us to visually examine the performance measures for different settings using the simulation-based optimization approach. If the performance measure under consideration has to be minimized (e.g., number of air taxis, cost per air taxi ride), the ideal value of each performance measure $m$ ($A_m$) is the minimum value of that performance measure observed across all the settings. The normalized value ($N_{m,s}$) is obtained by dividing $A_m$ by the value of performance measure for that setting $s$ ($P_{m,s}$), as given in Equation (12). On the other hand, if the performance measure has to be maximized (e.g., number of customers traveling per day per air taxi, utilization), $A_m$ will be the maximum value of that performance measure observed across all settings, and the normalized value ($N_{m,s}$) is obtained by dividing $P_{m,s}$ by $A_m$, as given in Equation (13).

If $m$ has to be minimized, the normalized value of $m$ for setting $s$, $N_{m,s} = \dfrac{A_m}{P_{m,s}}$ (12)

If $m$ has to be maximized, the normalized value of $m$ for setting $s$, $N_{m,s} = \dfrac{P_{m,s}}{A_m}$ (13)

Based on Equations (12) and (13), it is evident that $0 \leq N_{m,s} \leq 1$, and a higher value of $N_{m,s}$ indicates that the performance measure is closer to the ideal value, and hence, more preferable.

In the value path approach, the solution obtained from each setting is represented by a line, and all the measures can be visually compared using a single figure. If a line is above another line, then the solution obtained from the first setting is dominating the other (Cintron et al., 2010). Figure 10 clearly indicates that TS-1 has the ideal utilization rate with the least cost/passenger, nevertheless, the number of air taxis required is about 33% more than that of the base case instance. Whereas, TS-4 performs the best for the required number of vehicle metric, and is observed to be the worst for all the others.

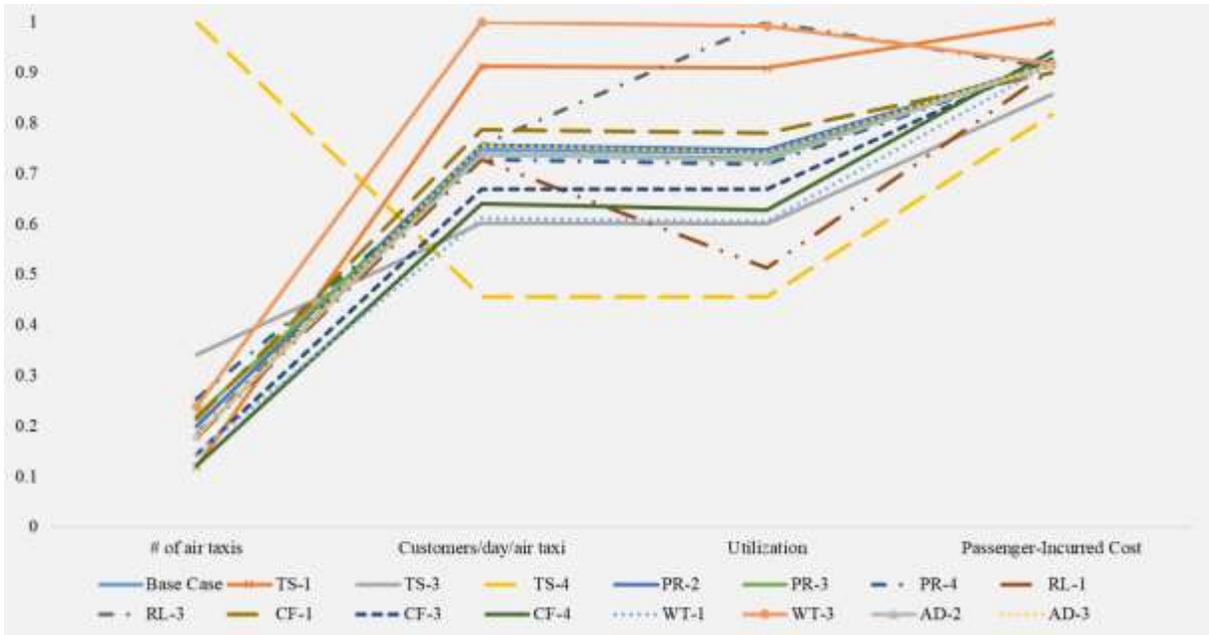

**Figure 10:** Value Path Graph of Different Settings

Based on the decision maker's estimates on the percentage of demand fulfillment or passenger's willingness to fly rate, he/she can decide on the required number of air taxis.

The following managerial implications are listed for various settings:

(1) Results indicate that the daily average number of customers served by each air taxi remains almost the same, except for the change in the time savings and customer wait time parameters. Both are found to be exponentially impacting the required number of air taxis, and hence, it is necessary to conduct an intense study to estimate these parameters.

(2) Although the on-road travel limit seems to have an insignificant impact on the number of air taxis or the average number of customers served, this parameter appears to be sensitive with respect to the number of infrastructures as well as the utilization. Hence, market analysis is strongly encouraged for this factor, as well.

(3) The aviation service appears to be utilized the highest when the maximum customer wait time is 15 minutes, with more than 20% decrease in the required number of air taxis from the base case. However, under this setting, from a customer's standpoint, it might appear beneficial only to avail of air taxi services for very long-distance travel, due to the substantial wait time.

(4) Only a small proportion of customers are traveling over 25 miles. Hence, in case if the logistics company is initially interested in starting limited air taxi operations, it is ideal for providing services to customers traveling between 10 and 25 miles.

(5) We see that 12 AM - 6 AM is a lean period, and similarly, the demand during weekends is noticeably lower than that of the weekdays. It is necessary for companies to consider these off-peak hours while developing pricing strategies to improve vehicle utilization.

(6) The demand between 3 and 6 PM is the highest, and it is strongly recommended to introduce ride-sharing services during these hours.

## 6. Conclusions

Presumed to launch in the forthcoming years, air taxis are compact aviation vehicles that will be operating in metropolitan cities with the goal of providing faster commutes to millions of passengers on a daily basis.

### 6.1 Paper Contributions

Most of the previous studies in the air taxi domain exclusively focus on vehicle design. To the best of the author's knowledge, this paper is the first to develop an algorithm to effectively dispatch and route air taxis in a cyber-physical system in real-time. The proposed simulation-based optimization approach will provide dynamic routing solutions for each air taxi, considering several unique constraints. The model developed in this paper can be used by any company that is interested in venturing into this air taxi market. The application of the proposed algorithm employs the data from a prior study containing millions of potential air taxi customer demand. Using the dispatching tool, the number of air taxis required, utilization rate, the daily number of customers served, and the passenger-incurred cost are reported for several parameters settings. The results of the proposed model are compared with those of the existing method in the literature, and superior results are achieved with the developed approach for numerous problem instances.

## 6.2 Managerial Implications

This paper also proposes the following managerial recommendations that are generic and can be used while implementing air taxi network operations in any metropolitan city:

- The time savings and customer wait time metrics exponentially influence certain performance measures, such as the daily average number of customers served per air taxi and the required number of air taxis, and hence, it is necessary to conduct an intense study to estimate these parameters.
- Another measure that requires market analysis is the on-road travel limit. Though this seems to have an insignificant impact on the number of air taxis or the average number of customers served, this parameter appears to be sensitive to the number of infrastructures as well as the utilization.
- When initially venturing into the market, it is recommended that services can be provided to customers traveling between 10 and 25 miles.
- Ride-sharing is strongly recommended to be introduced, particularly between 3 PM and 6 PM, when there is a very high demand for air taxis.

## 6.3 Limitations and Future Research

One of the main limitations of this research is the lack of data from other methods of transportation (such as subways and buses) for conducting the analysis. Nonetheless, as the proposed dispatching algorithm is generic, it can be used for any demand stream. This research assumes that vehicle maintenance and charging can be accommodated during the idle time of the air taxis. This assumption might be addressed in future studies by explicitly including these times while developing dispatching policies. As a future scope of improvement, in addition to flight scheduling, decisions for first- and last leg on-road car routing can be integrated into the algorithm. In the methodological end, one can develop a dynamic programming model for routing air taxis and compare the results of the proposed simulation-based optimization algorithm.

**Acknowledgment**

The author would like to thank undergraduate research assistant at the University of Missouri, Joshua Zack, for helping in this research endeavor.